\documentclass[runningheads,a4paper]{llncs}

\usepackage{cmap,llncsdoc}
\usepackage{amsmath,amsfonts} % Math packages
\usepackage[T1]{fontenc}
\DeclareMathOperator*{\argmax}{arg\,max}
\DeclareMathOperator*{\argmin}{arg\,min}
\usepackage{graphicx,multirow,subfigure,algorithm,algorithmicx,algpseudocode}

\usepackage[ngerman,english]{babel}
\addto\extrasenglish{\languageshorthands{ngerman}\useshorthands{"}}

\usepackage[%
rm={oldstyle=false,proportional=true},%
sf={oldstyle=false,proportional=true},%
tt={oldstyle=false,proportional=true,variable=true},%
qt=false%
]{cfr-lm}
\usepackage[math]{blindtext}

\usepackage{cite}

\usepackage{paralist}

\usepackage{csquotes}

\usepackage{microtype}

\usepackage{url}
\makeatletter
\g@addto@macro{\UrlBreaks}{\UrlOrds}
\makeatother
\usepackage{xcolor}

\usepackage{pdfcomment}
\hypersetup{hidelinks,
   colorlinks=true,
   allcolors=black,
   pdfstartview=Fit,
   breaklinks=true}
\usepackage[all]{hypcap}

\usepackage[capitalise,nameinlink]{cleveref}
\crefname{section}{Sect.}{Sect.}
\Crefname{section}{Section}{Sections}

\usepackage{xspace}
\usepackage{color}
\begin{document}

\title{Investigating Network for Fake News Study}
\title{Understanding and Modeling Fake News via Network Analysis}
% \title{Detecting and Mitigating Fake News: \\A Network Perspective}
\title{Studying Fake News via Network Analysis:\\Detection and Mitigation}

% \title{Understanding and Modeling Networks for Fake News Study}
% \title{TBD}
%If Title is too long, use \titlerunning
% \titlerunning{TBD}
\author{Kai Shu\inst{1}, H. Russell Bernard\inst{2} \and Huan Liu\inst{1}}
\institute{
Computer Science and Engineering
\and
Institute for Social Science Research\\ Arizona State University, Tempe, AZ, USA\\
\email{\{kai.shu, asuruss, huan.liu\}}@asu.edu
}
\maketitle

\begin{abstract}
Social media is becoming increasingly popular for news consumption due to its easy access, fast dissemination, and low cost. However, social media also enables the wide propagation of ``fake news'', i.e., news with intentionally false information.  Fake news on social media can have significant negative societal effects.  Identifying and mitigating fake news also presents unique challenges. To tackle these challenges, many existing research efforts exploit various features of the data, including network features. In essence, a news dissemination ecosystem involves three dimensions on social media, i.e., a content dimension, a social dimension, and a temporal dimension. In this chapter, we will review network properties for studying fake news, introduce popular network types and propose how these networks can be used to detect and mitigate fake news on social media.  
\end{abstract}


\section{Introduction}\label{sec:intro}
Social media has become an important means of large-scale information sharing and communication in all occupations, including marketing, journalism, public relations, and more~\cite{zafarani2014social}. The reasons for this change in consumption behaviors are clear: (i) it is often faster and cheaper to consume news on social media compared to news on traditional media, such as newspapers or television; and (ii) it is easier to share, comment on, and discuss the news with friends or other readers on social media. However, the low cost, easy access, and rapid dissemination of information of social media draws a large audience and enables the wide propagation of ``fake news'', i.e., news with intentionally false information. Fake news on social media is growing fast in volume and can have negative societal impacts. First, people may accept deliberate lies as truths~\cite{paul2016russian}; second, fake news can change the way people respond to legitimate news; and finally, the prevalence of fake news has the potential to break the trustworthiness of the entire news ecosystem. In this chapter, we discuss recent advancements--based on a network perspective--for the detection and mitigation of fake news.

Fake news on social media presents unique challenges. First, fake news is intentionally written to mislead readers, which makes it nontrivial to detect simply based on content. Second, social media data is large-scale, multi-modal, mostly user-generated, sometimes anonymous and noisy. Third, the consumers of social media come from different backgrounds, have disparate preferences or needs, and use social media for varied purposes. Finally, the low cost of creating social media accounts makes it easy to create malicious accounts, such as social bots, cyborg users and trolls, all of which can become powerful sources and proliferation of fake news.

\begin{figure}[htbp!]
   \centering
   \includegraphics[scale=0.7]{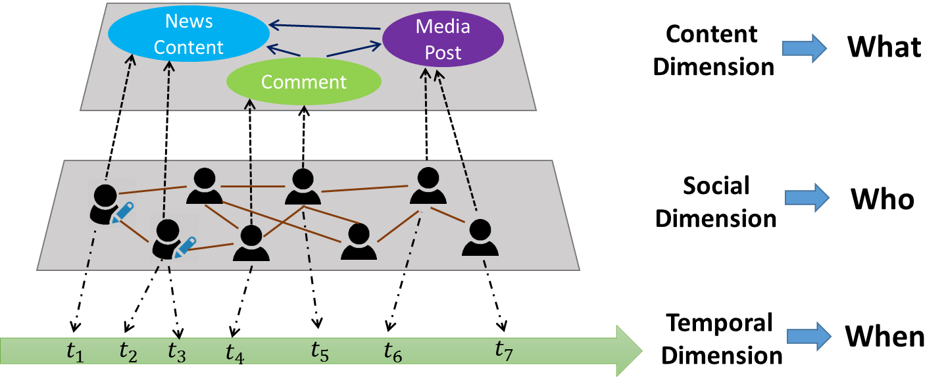}
   \caption{The Information Dimensions of News Dissemination Ecosystem.}
   \label{fig:layer}
\end{figure}

The news dissemination ecosystem on social media involves three dimensions (Figure~\ref{fig:layer}), a content dimension (``What''), a social dimension (``Who''), and a temporal dimension (``When''). The content dimension describes the correlation among news pieces, social media posts, comments, etc. The social dimension involves the relations among publishers, news spreaders and consumers. The temporal dimension illustrates the evolution of users' publishing and posting behaviors over time. As we will show, we can use these relations to detect and mitigate the effects of fake news.

% Motivation of network analysis is needed. fake news detection/mitigation
%Network analysis is shown to be important and useful in different social network tasks.  For fake news study, two important tasks are fake news detection and mitigation. Fake news detection can be formalized as a classification task that requires feature extraction and model construction. Recent advancements of network representation learning, such as network embedding and deep neural networks, open up the doors to better capture the features of news from auxiliary information such as friendship network, temporal user engagements, interaction networks. In addition, knowledge networks as auxiliary information can help evaluate the news veracity through network matching such as path finding and flow optimization. Fake news mitigation aims to restrict the spread of fake news by proactively blocking target users or starting a mitigating campaign at an early stage. First, network diffusion models can be applied to find fake news provenance nodes and provenance paths. Second, the impact of fake news can be assessed and mitigated through network estimation and network influence minimization strategies.  Thus, it's necessary to explore network analysis for fake news detection and mitigation.

\textit{Detection} of fake news can be formalized as a classification task that requires feature extraction and model construction. Recent advancements of network representation learning, such as network embedding and deep neural networks, allow us to better capture the features of news from auxiliary information such as friendship network, temporal user engagements, and interaction networks. In addition, knowledge networks as auxiliary information can help evaluate the veracity of news through network matching operations such as path finding and flow optimization. For \textit{mitigation}, the aim is to proactively block target users or start a mitigating campaign at an early stage. We will show that network diffusion models can be applied to trace the provenance nodes and provenance paths of fake news. In addition, the impact of fake news can be assessed and mitigated through network estimation and network influence minimization strategies. We begin, then, with an introduction to network properties.

%In this chapter, we will focus on network analysis for fake news study. The rest of this chapter is organized as follows. In Section~\ref{sec:net_property}, we will introduce the background of fake news ecosystem with several network properties. We will demonstrate different types of networks that are commonly constructed to study fake news in Section~\ref{sec:net_type}. In Section~\ref{sec:detection}, we introduce how features and models can be extracted and constructed to detect fake news from the perspective of network analysis from different types of networks. We further describe how to mitigate fake news effects with network estimation and intervention strategies in Section~\ref{sec:mitigation}. Finally, we conclude this chapter in Section~\ref{sec:discuss}.

\section{Network Properties} \label{sec:net_property}

In this section, we outline the potential role of network properties for the study of fake news. First, users form groups with like-minded people, resulting in what are widely known as \textit{echo chambers}. Second, \textit{individual users} play different roles in the dissemination of fake news. Third, social media platforms allow users to personalize how information is presented to them, thus isolating users from information outside their personalized \textit{filter bubbles}. Finally, highly active \textit{malicious user accounts} become powerful sources and proliferators of fake news.

\subsection{Echo Chambers}

The process of seeking and consuming information on social media is becoming less mediated. Users on social media tend to follow like-minded people and thus receive news that promotes their preferred, existing narratives. This may increase social polarization, resulting in an \textit{echo chamber} effect~\cite{barbera2015tweeting}. The echo chamber effect facilitates the process by which people consume and believe fake news based on the following psychological factors~\cite{paul2016russian}: i) \textit{social credibility}, which means people are more likely to perceive a source as credible if others perceive it as such, especially when there is not enough information available to assess the truthfulness of that source; and ii) \textit{frequency heuristic}, which means that consumers may naturally favor information they hear frequently, even if it is fake news. In echo chambers, users share and consume the same information, which creates segmented and polarized communities.

\subsection{Individual Users}

During the fake news dissemination process, individual users play different roles. For example, i) \textit{persuaders} spread fake news with supporting opinions to persuade and influence others to believe it; ii) \textit{gullible users} are credulous and easily persuaded to believe fake news; and iii) \textit{clarifiers} propose skeptical and opposing viewpoints to clarify fake news. Social identity theory~~\cite{tajfel2004social} suggests that social acceptance and affirmation is essential to a person's identity and self-esteem, making persuaders likely to choose ``socially safe'' options when consuming and disseminating news information. They follow the norms established in the community even if the news being shared is fake news. The cascade of fake news is driven not only by influential persuaders but also by a critical mass of easily influenced individuals~\cite{del2016echo}, i.e., gullible users. Gullibility is a different concept from trust. In psychological theory~\cite{rotter1980interpersonal}, general trust is defined as the default expectations of other people's trustworthiness. High trusters are individuals who assume that people are trustworthy unless proven otherwise. Gullibility, on the other hand, is insensitivity to information revealing untrustworthiness. Reducing the diffusion of fake news to gullible users is critical to mitigating fake news. Clarifiers can spread opposing opinions against fake news and avoid one-sided viewpoints. Clarifiers can also spread true news which can: i) immunize users against changing  beliefs before they are affected by fake news; and ii) further propagate and spread true news to other users.

\subsection{Filter Bubbles}

A filter bubble is an intellectual isolation that occurs when social media websites use algorithms to personalize the information a user would want to see~\cite{pariser2011filter}. The algorithms make assumptions about user preferences based on the user's historical data, such as former click behavior, browsing history, search history, and location. Given these assumptions, the website is more likely to present information that will support the user's past online activities. A filter bubble can reduce connections with contradicting viewpoints, causing the user to become intellectually isolated. A filter bubble will amplify the individual psychological challenges to dispelling fake news. These challenges include: i) \textit{Na\"{i}ve Realism}~\cite{ward1997naive}: consumers tend to believe that their perceptions of reality are the only accurate views, while others who disagree are regarded as uninformed, irrational, or biased; and ii) \textit{Confirmation Bias}~\cite{nickerson1998confirmation}: consumers prefer to receive information that confirms their existing views.

\subsection{Malicious Accounts}

Social media users can be malicious, and some malicious users may not even be real humans. Malicious accounts that can amplify the spread of fake news include social bots, trolls, and cyborg users. Social bots are social media accounts that are controlled by a computer algorithm. The algorithm automatically produces content and interacts with humans (or other bot users) on social media. Social bots can be malicious entities designed specifically for manipulating and spreading fake news on social media. Trolls are real human users who aim to disrupt online communities and provoke consumers to an emotional response. Trolls enable the easy dissemination of fake news among otherwise normal online communities. Finally,  cyborg users can spread fake news in a way that blends automated activities with human input. Cyborg accounts are usually registered by a human as a disguise for automated programs that are set to perform activities on social media. The easy switch between humans and bots offer cyborg users unique opportunities to spread fake news. 

\section{Network Types} \label{sec:net_type}

In this section, we introduce several network structures that are commonly used to detect and mitigate fake news. Then, following the three dimensions of the news dissemination ecosystem outlined above, we illustrate how \textit{homogeneous} and \textit{heterogeneous} networks can be built within a specific dimension and  across dimensions.

\subsection{Homogeneous Networks}
\begin{figure}[htbp!]
\centering
\subfigure[Friendship Network]{
  {\includegraphics[scale=0.47]{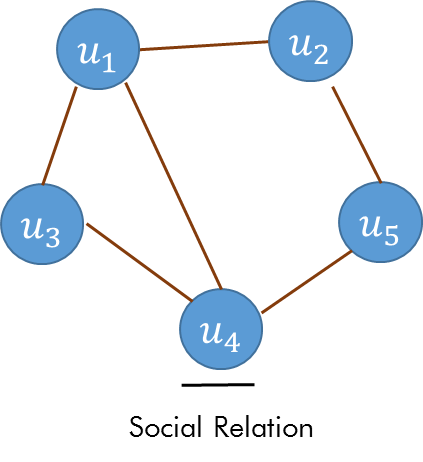}}
}
\hspace{-0.1cm}
\subfigure[Diffusion Network]{
  {\includegraphics[scale=0.47]{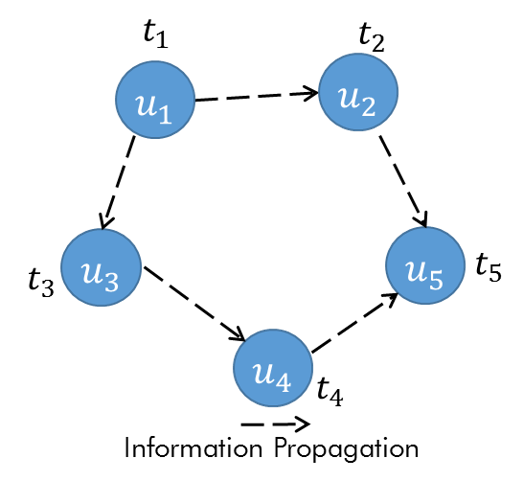}}
}
\hspace{-0.1cm}
\subfigure[Credibility Network]{
  {\includegraphics[scale=0.47]{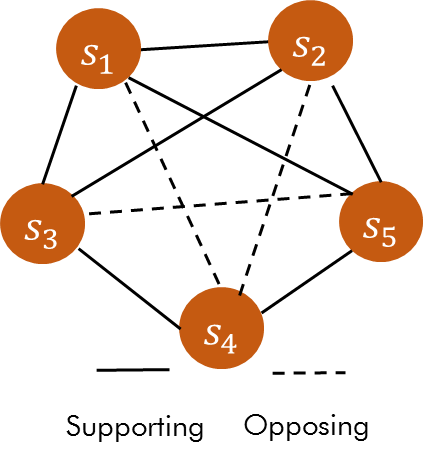}}
}
\caption{Homogeneous Networks. Three types of homogeneous networks are illustrated: friendship network, diffusion network, and credibility network. Node $u$ indicates a user, and $s$ represents a social media post.}\label{fig:homo}
\end{figure}

Homogeneous networks have the same node and link types. As shown in Figure~\ref{fig:homo}, we introduce three types of homogeneous networks: friendship networks, diffusion networks, and credibility networks. Each of these types is potentially useful in detecting and mitigating fake news. 
% The motivation of why these networks could help is missing.
\\\\
\textbf{Friendship Networks} A user's friendship network in the social layer can be represented as a directed graph $G_F=(U, E_F)$, where $U$ and $E_F$ are the node and edge sets, respectively. A node $u \in U$ represents a user, and $(u_1,u_2)\in E$ represents whether a social relation exists.

Homophily theory~\cite{mcpherson2001birds} suggests that users tend to form relationships with like-minded friends, rather than with users who have opposing preferences and interests. Likewise, social influence theory~\cite{marsden1993network} predicts that users are more likely to share similar latent interests towards news pieces. Thus, the friendship network provides the structure to understand the set of social relationships among users. The friendship network is the basic route for news spreading and can reveal community information.
\\\\
\textbf{Diffusion Networks} A diffusion network in the social layer can be represented as a directed graph $G_D=(U, E_D, p, t)$, where $U$ and $E$ are the node and edge sets, respectively. A node $u \in U$ represents an entity, which can publish, receive, and propagate information at time $t_i\in t$. A directed edge, $(u_1 \rightarrow u_2) \in E_D$, between nodes $u_1, u_2 \in U$ represents the direction of information propagation. Each directed edge $(u_1 \rightarrow u_2) \in E_D$, between nodes $u_1, u_2 \in U$ represents the direction of information propagation. Each directed edge $(u_1 \rightarrow u_2)$ is assumed to be associated with an information propagation probability, $p(u_1 \rightarrow u_2) \in [0, 1]$.

The diffusion network is important for learning about representations of the structure and temporal patterns to help identify fake news. By discovering the sources of fake news and the spreading paths among the users, we can also better mitigate fake news problem.
\\\\
\textbf{Credibility Networks} A credibility network~\cite{jin2016news} can be represented as an undirected graph $G_C=(V,E_C,s)$, where $V$ denotes the set of social media posts with corresponding credibility scores that are related to original news pieces, and the edges $E$ denote the link type, such as supporting and opposing, between two nodes. and $c(v_1,v_2)$ indicates the  conflicting degree of credibility values of node $v_1$ and $v_2$.

Users express their viewpoints toward original news pieces through social media posts. In these posts, they can either express the same viewpoints (which mutually support each other), or conflicting viewpoints (which may reduce their credibility scores). By modeling these relationships, the credibility network can be used to evaluate the overall truthfulness of news by leveraging the credibility scores of each social media post relevant to the news.

\subsection{Heterogeneous Networks}

\begin{figure}[htbp!]
\centering
\subfigure[Knowledge Network]{
  {\includegraphics[scale=0.47]{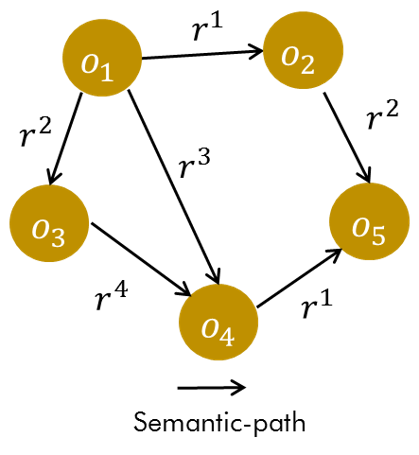}}
}\label{fig:knowledge}
\hspace{0.1cm}
\subfigure[Stance Network]{
  {\includegraphics[scale=0.47]{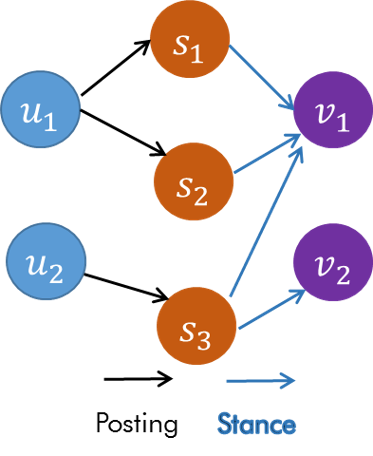}}
}\label{fig:stance}
\hspace{0.1cm}
\subfigure[Interaction Network]{
  {\includegraphics[scale=0.47]{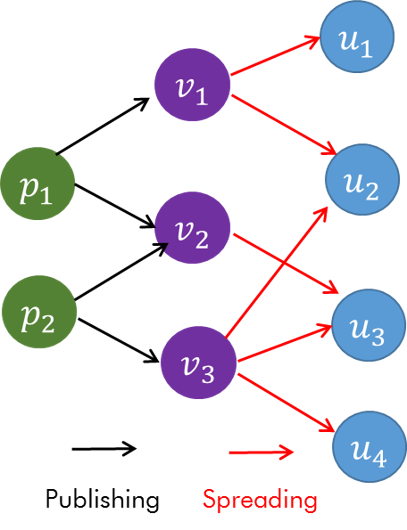}}
}\label{fig:interaction}
\caption{Heterogeneous Networks. Three types of homogeneous networks are illustrated: knowledge network, stance network, and interaction network. Node $o$ indicates a knowledge entity, $v$ represents a news item, and $p$ means a news publisher.}\label{fig:hete}
\vspace{-0.2cm}
\end{figure}
%Heterogeneous networks refer to those networks that have the different node types and link types. The advantages of heterogeneous networks are the abilities to represent and encode information and relations from different aspects. During the news dissemination process, there are different types of entities involved, e.g., users, social media posts, news, etc. As shown in Figure~\ref{fig:hete}, these are the common types of heterogeneous networks for fake news analysis, detail described as below.

Heterogeneous networks have a different set of node and link types. The advantages of heterogeneous networks are the abilities to represent and encode information and relationships from different perspectives. During the news dissemination process, different types of entities are involved, including users, the social media posts, the actual news, etc. Figure~\ref{fig:hete} shows the common types of heterogeneous networks for analyzing fake news: knowledge networks, stance networks, and interaction networks.
\\\\
\textbf{Knowledge Networks} A knowledge network (i.e., knowledge graph) $G_K=(I,E_I,R)$ is constructed with the nodes $I$ representing the knowledge entities, edges $E_I$ indicating the relation between them, $R$ is the relation sets, and $g:E\rightarrow R$ is the function labeling each edge with a semantic predicate.

The knowledge network integrates linked open data, such as DBdata and Google Relation Extraction Corpus (GREC), as a heterogeneous network topology. Fact-checking using a knowledge graph  checks whether the claims in news content can be inferred from existing facts in the knowledge networks~\cite{ciampaglia2015computational,shi2016fact}.
\\\\
\textbf{Stance Networks} A heterogeneous stance network can be represented as a heterogeneous network $G_S=(\{U,S,V\},E_S)$, where the nodes can be users, news items and social media posts; and the edges $E_S$ denote the link type between two nodes, such as \textit{posting} between users and posts, and \textit{stance} between posts and news items. The stances are treated as important signals and can be aggregated to infer the news veracity.

Stances (or viewpoints) indicate the users' opinions towards the news, such as supporting, opposing, etc. Typically, fake news pieces will provoke tremendous controversial views among social media users, in which denying and questioning stances are found to play a crucial role in signaling claims as being fake.
\\\\
\textbf{Interaction Networks} An interaction network $G_I=(\{P,U,V\},E_I)$ consists of nodes representing publishers, users, news, and the edges $E_I$ indicating the interactions among them. For example, edge $(p\rightarrow v)$ demonstrates that publisher $p$ publishes news item $v$, and $(v\rightarrow u)$ represents news $v$ is spread by user $u$.

The interaction networks can represent the correlations among different types of entities, such as publisher, news, and social media post,  during the news dissemination process~\cite{shu2017exploiting}. The characteristics of publishers and users, and the publisher-news and news-users interactions have potential to differentiate fake news. 

% \cite{jin2014news}

\section{Fake News Detection} \label{sec:detection}

%Fake news detection evaluates the truthful value of a news piece, which can be formalized as a classification problem. The common procedure is feature extraction and model construction. In feature extraction, we  capture the differentiable characteristics of news pieces to construct effective representations; Based on these representations, we can construct various models to learn and transform the features into a predicted label. To this end, we introduce how features and models can be extracted and constructed through network analysis on different types of networks.

Fake news detection evaluates the truth value of a news piece, which can be formalized as a classification problem. The common procedure is feature extraction and model construction. In feature extraction, we capture the differentiable characteristics of news pieces to construct effective representations; Based on these representations, we can construct various models to learn and transform the features into a predicted label. To this end, we introduce how features and models can be extracted and constructed in different types of networks.

\subsection{Interaction Network Embedding}
Interaction networks describe the relationships among different entities such as publishers, news pieces, and users. Given the interaction networks the goal is to embed the different types of entities into the same latent space, by modeling the interactions among them. We can leverage the resultant feature representations of news to perform fake news detection.

\textbf{News Embedding} We can use news content to find clues to differentiate fake news and true news. Using non-negative Matrix Factorization (NMF) we can attempt to project the document-word matrix to a joint latent semantic factor space with low dimensionality, such that the document-word relations are modeled as the inner product in the space. Specifically, giving the news-word matrix $\mathbf{X}\in\mathbb{R}_{+}^{n \times t}$, NMF methods try to find two nonnegative matrices $\mathbf{D}\in \mathbb{R}_{+}^{n\times d}$ and $\mathbf{V}\in \mathbb{R}_{+}^{t\times d}$ by solving the following optimization problem,

\begin{equation}
\begin{aligned}
  \min_{\substack{\mathbf{D} ,\mathbf{V}\geq0}} & \|\
  \mathbf{X} - \mathbf{D}\mathbf{V}^T\|_F^2 \\
\end{aligned}\label{eqn:news}
\end{equation}
where $d$ is the dimension of the latent topic space. In addition, $\mathbf{D}$ and $\mathbf{V}$ are the nonnegative matrices indicating low-dimensional representations of news and words. 
%The term $\lambda (\|\mathbf{D}\|_F^2+ \|\mathbf{V}\|_F^2)$ is introduced to avoid over-fitting.

\textbf{User Embedding.} On social media, people tend to form  relationships with like-minded friends, rather than with users who have opposing preferences and interests~\cite{shu2018understand}. Thus, connected users are more likely to share similar latent interests in news pieces. To obtain a standardized representation, we use nonnegative matrix factorization to learn the user's latent representations (we will introduce other methods in Section~\ref{sec:friend}). Specifically, giving user-user adjacency matrix $\mathbf{A}\in\{0,1\}^{m\times m}$, we learn nonnegative matrix $\mathbf{U}\in\mathbb{R}_{+}^{m\times d}$ by solving the following optimization problem,
\begin{equation}
\begin{aligned}
  \min_{\substack{\mathbf{U},\mathbf{T}\geq 0}} & \|\mathbf{Y}\odot(\mathbf{A} - \mathbf{U} \mathbf{T}\mathbf{U}^T)\|_F^2
\end{aligned}
\end{equation}
where $\mathbf{U}$ is the user latent matrix, $\mathbf{T}\in\mathbb{R}_{+}^{d \times d}$ is the user-user correlation matrix, and $\mathbf{Y}\in\mathbb{R}^{m \times m}$ controls the contribution of $\mathbf{A}$. Since only positive samples are given in $\mathbf{A}$, we can first set $\mathbf{Y}=sign(\mathbf{A})$, then perform negative sampling and generate the same number of unobserved links and set weights as 0. 

\textbf{User-News Embedding} The user-news interactions can be modeled by considering the relationships between user attributes and the level of veracity of news items. Intuitively, users with low credibilities are more likely to spread fake news, while users with high credibility scores are less likely to spread fake news. Each user has a credibility score that we can infer using his/her published posts~\cite{abbasi2013measuring}, and we use $\mathbf{c} = \{\mathbf{c}_1,\mathbf{c}_2,...,\mathbf{c}_m\}$ to denote the credibility score vector, where a larger $c_i\in[0,1]$ indicates that user $u_i$ has a higher credibility. The user-news engaging matrix is represented as $\mathbf{W}\in\{0,1\}^{m\times n}$, where $\mathbf{W}_{ij} = 1$ indicates that user $u_i$ has engaged in the spreading process of the news piece $v_j$ ; otherwise $\mathbf{W}_{ij} = 0$. The user-news embedding objective function is shown as follows,
\begin{equation}
\begin{aligned}
\min~\underbrace{\sum_{i=1}^m \sum_{j=1}^r \mathbf{W}_{ij} \mathbf{c}_i  (1 - \frac{1+\mathbf{y}_{Lj}}{2}) ||\mathbf{U}_i - \mathbf{D}_{j}||_2^2 }_{\text{True news}}\\+\underbrace{\sum_{i=1}^m \sum_{j=1}^r  \mathbf{W}_{ij} (1-\mathbf{c}_i) (\frac{1+\mathbf{y}_{Lj}}{2}) ||\mathbf{U}_i - \mathbf{D}_{j}||_2^2}_{\text{Fake news}}
\end{aligned} \label{eqn_user_engage}
\end{equation}
where $\mathbf{y}_L\in\mathbb{R}^{r \times 1}$ is the label vector of all partially labeled news. The objective considers two situations: i) for true news, i.e., $\mathbf{y}_{Lj}=-1$, which ensures that the distance between latent features of high-credibility users and that of true news is small; and ii) for fake news, i.e., $\mathbf{y}_{Lj}=1$, which ensures that the distance between the latent features of low-credibility users and that of true news is small.

\textbf{Publisher-News Embedding} The publisher-news interactions can be modeled by incorporating the characteristics of the publisher and news veracity values. Fake news is often written to convey opinions or claims that support the partisan bias of the news publisher. Publishers with a high degree of political bias are more likely to publish fake news~\cite{shu2017exploiting}. Thus, a useful news representation should be good at predicting the partisan bias score of its publisher. The partisan bias scores are collected from fact-checking websites and can be represented as a vector $\mathbf{o}$. We can utilize publisher partisan labels vector $\mathbf{o}\in \mathbb{R}^{l\times 1}$ and publisher-news matrix $\mathbf{B}\in\mathbb{R}^{l \times n}$ to optimize the news feature representation learning as follows,

\begin{equation}
\begin{aligned}
  \min~ \|\
  \bar{\mathbf{B}}\mathbf{D}\mathbf{Q}-\mathbf{o}\|_2^2
\end{aligned}
\end{equation}
where the latent feature of news publisher can be represented by the features of all the news it published, i.e., $\mathbf{\bar{\mathbf{B}}D}$. $\bar{\mathbf{B}}$ is the normalized user-news publishing relation matrix, i.e., $\bar{\mathbf{B}}_{kj} = \frac{\mathbf{B}_{kj}}{\sum_{j=1}^n\mathbf{B}_{kj}}$. $\mathbf{Q}\in \mathbb{R}^{d\times 1}$ is the weighting matrix that maps news publishers' latent features to corresponding partisan label vector $\mathbf{o}$.

The finalized model combines all previous components into a coherent model. In this way, we can obtain the latent representations of news items $\mathbf{D}$ and of users $\mathbf{U}$ through the network embedding procedure, which can be utilized to perform fake news classification tasks. 

\subsection{Temporal Diffusion Representation}
\begin{figure}[htbp!]
   \centering
   \includegraphics[scale=0.49]{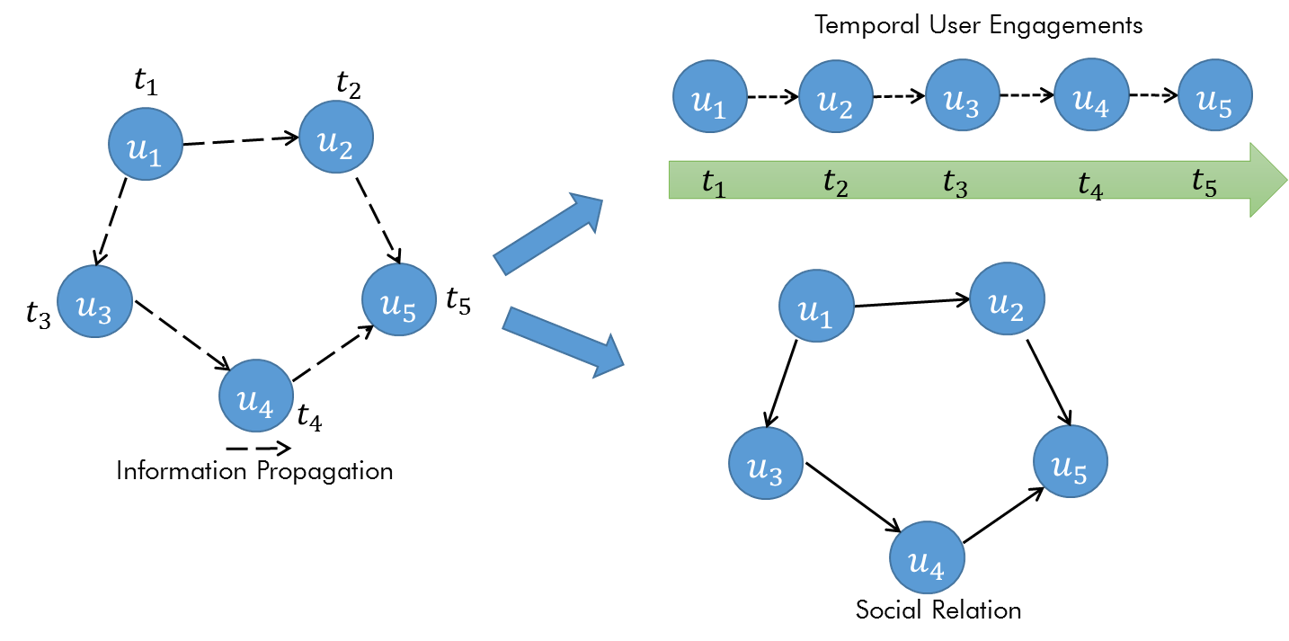}
   \caption{A diffusion network consists of temporal user engagements and a friendship network.}
   \label{fig:temporal}
\end{figure}
The news diffusion process involves abundant temporal user engagements on social media~\cite{ruchansky2017csi,shu2017fake,wu2018tracing}. The social news engagements can be defined as a set of tuples $E = {e_i}$ to represent the process of how news items spread over time among $m$ users in $U = \{u_1, u_2, ..., u_m\}$. Each engagement $e_i = \{u_i, t_i, s_i\}$ represents that a user $u_i$ spreads news article at time $t_i$ by posting $s_i$. As shown in Figure~\ref{fig:temporal}, the information diffusion network consists of two major parts of knowledge: i) temporal user engagements and ii) a friendship network. For example, a diffusion path between two users $u_i$ and $u_j$ exists if and only if (1) $u_j$ follows $u_i$; and (2) $u_j$ posts about a given news only after $u_i$ does so. 

The goal of learning temporal representations is to capture the user's pattern of temporal engagements with a news article $v_j$. Recent advances in the study of deep neural networks, such as Recurrent Neural Networks (RNN), have shown promising performance for learning representations. RNN are powerful structures that allow the use of loops within the neural network to model sequential data.  Given the diffusion network $G_D$, the key procedure is to construct meaningful features $\mathbf{x}_i$ for each engagement $e_i$. The features can generally be extracted from the contents of $s_i$ and the attributes of $u_i$. For example, $\mathbf{x}_i$ consists of the following components: $\mathbf{x}_i = (\eta, \Delta t, \mathbf{x}_{u_i}, x_{s_i})$. The first two variables $\eta$ and $\Delta t$, represent the number of total user engagements through time $t$ and the time difference between engagements, respectively. These variables capture the general measure of frequency and time interval distribution of user engagements of the news piece $v_j$. For the content features of users posts, the $\mathbf{x}_{s_i}$ can be extracted from hand-crafted linguistic features, such as n-gram features, or by using word embedding methods such as doc2vec~\cite{le2014distributed} or GloVe~\cite{pennington2014glove}. We can extract the features of users $\mathbf{x}_{u_i}$ by performing a singular value decomposition of the user-news interaction matrix $\mathbf{W}\in\{0,1\}^{m\times n}$, where $\mathbf{W}_{ij} = 1$ indicate that user $u_i$ has engaged in the process of spreading the news piece $v_j$ ; otherwise $\mathbf{W}_{ij} = 0$.

\begin{figure}[htbp!]
   \centering
   \includegraphics[scale=0.49]{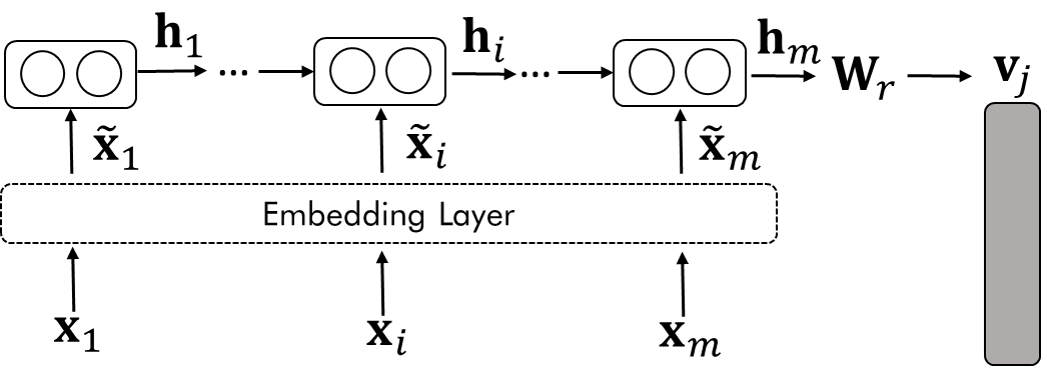}
   \caption{RNN framework for learning news temporal representations.}
   \label{fig:rnn}
\end{figure}

The RNN framework for learning news temporal representations is demonstrated in Figure~\ref{fig:rnn}. Since $\mathbf{x}_i$ includes features that come from different information space, such as temporal and content features, so we do not suggest incorporating $\mathbf{x}_i$ into RNN as the raw input. Thus, we can add a fully connected embedding layer to convert the raw input $\mathbf{x}_i$ into a standardized input features $\mathbf{\tilde{x}}_i$, in which the parameters are shared among all raw input features $\mathbf{x}_i, i=1,...,m$. Thus, the RNN takes a sequence $\mathbf{\tilde{x}}_1,\mathbf{\tilde{x}}_2,...,\mathbf{\tilde{x}}_m$ as input. At each time-step $i$, the output of previous step $\mathbf{h}_{i-1}$, and the next feature input $\mathbf{\tilde{x}}_i$ are used to update the hidden state $\mathbf{h}_i$. The hidden states $\mathbf{h}_i$ is the feature representation of the sequence up to time $i$ for the input engagement sequence. Thus, the hidden states of final step $\mathbf{h}_m$ is passed through a fully connected layer to learn the resultant news representation, defined as $\mathbf{v}_j = \text{tanh}(\mathbf{W}_r\mathbf{h}_m+\mathbf{b}_r)$, where $\mathbf{W}_r$ is the weight matrix and $\mathbf{b}_r$ is a bias vector.  Thus, we can use $\mathbf{v}_j$ to perform fake news detection and related tasks~\cite{ruchansky2017csi}.

\subsection{Friendship Network Embedding} \label{sec:friend}
News temporal representations can capture the evolving patterns of news spreading sequences. However, we lose the direct dependencies of users, which plays an important role in fake news diffusion. The fact that users are likely to form echo chambers, strengthens our need to model user social representations and to explore its added value for a fake news study. Essentially, given the friendship network $G_F$, we want to learn latent representations of users while preserving the structural properties of the network, including first-order and higher-order structure, such as second-order structure and community structure. For example, Deepwalk~\cite{perozzi2014deepwalk} can preserve the neighborhood structure of nodes by modeling a stream of random walks. In addition, LINE~\cite{tang2015line} can preserve both first-order and second-order proximities. Specifically, we can measure the first-order proximity by the joint probability distribution between the user $u_i$ and $u_j$,

\begin{equation}
p_1(u_i,u_j) = \frac{1}{1+exp(-\mathbf{u_i}^T\mathbf{u}_j)}
\end{equation}
where $u_i$ ($u_j$) is the social representation of user $u_i$ ($u_j$). We can model the second-order proximity by the probability of the context user $u_j$ being generated by the user $u_i$, as follows,

\begin{equation}
p_2(u_j|u_i) = \frac{exp(\mathbf{u_j}^T\mathbf{u}_i)}{\sum_{k=1}^{|V|} exp(\mathbf{u_k}^T\mathbf{u}_i)}
\end{equation}
where $|V|$ is the number of nodes or ``contexts'' for user $u_i$. This conditional distribution implies that users with similar distributions over the contexts are similar to each other. The learning objective is to minimize the KL-divergence of the two distributions and empirical distributions respectively. 

Network communities may actually be the more important structural dimension because fake news spreaders are likely to form polarized groups~\cite{shu2017fake}. This requires the representation learning methods to be able to model community structures. For example, a community-preserving node representation learning method, Modularized Nonnegative
Matrix Factorization  (MNMF), is proposed~\cite{wang2017community}. The overall objective is defined as follows,

\begin{equation}
\begin{aligned}
  \min_{\substack{\mathbf{M} ,\mathbf{U},\mathbf{H},\mathbf{C}\geq0}} & \underbrace{\|\
  \mathbf{S} - \mathbf{M}\mathbf{U}^T\|_F^2}_{\text{Proximity Mapping}} +\underbrace{\alpha \|\mathbf{H}-\mathbf{U}\mathbf{C}^T\|_F^2}_{\text{Community Mapping}}-\underbrace{\beta tr(\mathbf{H}^T\mathbf{B}\mathbf{H})}_{\text{Modularity Modeling}}\\
  &s.t. ~~~tr(\mathbf{H}^T\mathbf{H})=m
\end{aligned}\label{eqn:mnmf}
\end{equation} 
and comprises three major parts: proximity mapping, community mapping, and modularity modeling. In proximity mapping, $\mathbf{S}\in\mathbb{R}^{m\times m}$ is the user similarity matrix constructed from the user adjacency matrix (first-order proximity) and neighborhood similarity matrix (second-order proximity), and $\mathbf{M}\in\mathbb{R}^{m\times k}$ and $\mathbf{U}\in\mathbb{R}^{m\times k}$ are the basis matrix and user representations. For community mapping, $\mathbf{H}\in\mathbf{R}^{m\times l}$ is the user-community indicator matrix that we optimize to be reconstructed by the product of the user latent matrix $\mathbf{U}$ and the community latent matrix $\mathbf{C}\in\mathbf{R}^{l\times m}$. For modularity modeling, it represents the objective to maximize the modularity function~\cite{newman2006finding}, and $\mathbf{B}\in \mathbb{R}^{m \times m}$ is the modularity matrix.

\subsubsection{Credibility Network Propagation}
The basic assumption is that the credibility of a given news event is highly related to the credibilities of its relevant social media posts~\cite{jin2016news}. To classify whether a news item is true of fake, we can collect all relevant social media posts. Then, we can evaluate the news veracity score by averaging the credibility scores of all the posts.

%Credibility propagation approaches for fake news detection reason about the interrelations of relevant social media posts to predict news credibility. 

Given the credibility network $G_C$ for specific news pieces, the goal is to optimize the credibility values of each node (i.e., social media post), and infer the credibility value of corresponding news items~\cite{jin2016news}. In the credibility network $G_C$, there are i) a post credibility vector $\mathbf{T}=\{C(s_1),C(s_2),...,C(s_n)\}$ with $C(s_i)$ denoting the credibility value of post $s_i$; and ii) a matrix $\mathbf{W}\in\mathbb{R}^{n\times n}$, where $\mathbf{W}_{ij}=f(s_i,s_j)$ which denotes the viewpoint correlations between post $s_i$ and $s_j$, that is, whether the two posts take supporting or opposing positions.

\textbf{Network Initialization} Network initialization consists of two parts: node initialization and link initialization. First, we obtain the initial credibility score vector of nodes $\mathbf{T}_0$ from pre-trained classifiers with features extracted from external training data. The link is defined by mining the viewpoint relations, which are the relations between each pair of viewpoint such as contradicting or same. The basic idea is that posts with same viewpoints form supporting relations which raise their credibilities, and posts with contradicting viewpoints form  opposing relations which weaken their credibilities. Specifically, a social media post $s_i$ is modeled as a multinomial distribution $\theta_i$ over $K$ topics, and a topic k is modeled as a multinomial distribution $\psi_{tk}$ over $L$ viewpoints. The probability of a post $s_t$ over topic $k$ along with $L$ viewpoints is denoted as $p_{ik}=\theta_i\times\psi_{ik}$. The distance between two posts $s_t$ and $s_j$ are measured by using the Jensen-Shannon Distance: $Dis(s_i,s_j)=D_{JS}(p_{ik}||p_{jk})$.

The supporting or opposing relation indicator is determined as follows: it's assumed that one post contains a major topic-viewpoint, which can be defined as the largest proportion of $p_ik$. If the major topic-viewpoints of two posts $s_i$ and $s_j$ are clustered together (they take the same viewpoint), then they are mutually supporting; otherwise, they are mutually opposing. The similarity/dissimilarity measure of two posts are defined as:

\begin{equation}
\begin{aligned}
f(s_i,s_j) = \frac{(-1)^{a}}{D_{JS}(p_{ik}||p_{jk})+1}
\end{aligned}\label{eqn:link}
\end{equation}
where $a$ is the link type indicator, and if $a=0$, then $s_i$ and $s_j$ take the same viewpoint; otherwise, $a=1$.

%each post can be modeled as a pair of dependent mixtures: a mixture of topics and a mixture of viewpoints for each topic; and each topic-viewpoint pair is represented by a probability distribution over vocabulary terms. These topic viewpoints are then clustered into conflicting viewpoints clusters.

\textbf{Network Optimization} Posts with supporting relations should have similar credibility values; posts with opposing relations should have opposing credibility values. Therefore, the objective can be defined as a network optimization problem as below:

\begin{equation}
\begin{aligned}
  Q(\mathbf{T}) =& \mu\sum_{i,j=1}^{n}|\mathbf{W_{ij}}|\bigl(\frac{C(s_i)}{\sqrt{\bar{\mathbf{D}}_{ii}}}-b_{ij}\frac{C(s_j)}{\sqrt{\bar{\mathbf{D}}_{jj}}}\bigr)^2
  \\&+(1-\mu)\|\mathbf{T}-\mathbf{T}_0\|^2
\end{aligned}\label{eqn:propagation}
\end{equation}
where $\bar{\mathbf{D}}$ is a diagonal matrix with $\bar{\mathbf{D}}_{ii}=\sum_{k}|\mathbf{W}_{ik}|$ and $b_{ij} =1$, if $\mathbf{W}_{ij}\geq0$; otherwise $b_{ij} =0$. The first component is the smoothness constraint which guarantees the two assumptions of supporting and opposing relations; the second component is the fitting constraint to ensure variables not change too much from their
initial values; and $\mu$ is the regularization parameter to trade off two constraints. Then the credibility propagation on the proposed network $G_C$ is formulated as the minimization of this loss function:
\begin{equation}
\mathbf{T}^* = \argmin_{\mathbf{T}}Q(\mathbf{T})
\end{equation}
The optimum solution can be solved by updating $\mathbf{T}$ in an iterative manner through the transition function $\mathbf{T}(t)= \mu \mathbf{H}\mathbf{T}(t-1)+(1-\mu \mathbf{T}_0)$, where $\mathbf{H} = \bar{\mathbf{D}}^{-1/2}\mathbf{W}\bar{\mathbf{D}}^{-1/2}$. As the iteration converges, each post receives a final credibility value, and the average of them is served as the final credibility evaluation result for the news.

\subsection{Knowledge Network Matching}
In this section, we focus on exploiting knowledge networks to detect fake news. Knowledge networks are used as an auxiliary source to fact-check news claims. The goal is to match news claims with the facts in represented in knowledge networks.

\subsubsection{Path Finding}
Fake news spreads false claims in news content, so a natural means of detecting fake news is to check the truthfulness of major claims in the news article. Fact-checking methods use external sources such as knowledge networks, to assess the truthfulness of information. Specifically, a news claim can be checked automatically by finding the matching path to knowledge networks. A claim in news content can be represented by a subject-predicate-object triple $c=(s,p,o)$, where the subject entity $s$ is related to the object entity $o$ by the predicate relation $p$. 

We can find all the paths that start with $s$ and end with $o$, and then evaluate these paths to estimate the truth value of the claim. This set of paths, also named knowledge stream~\cite{shiralkar2017finding}, are denoted as $\mathcal{P}(s,o)$. Intuitively, if the paths involve more specific entities, then the claim is more likely to be true. Thus, we can define a ``specificity'' measure $S(P_{s,o})$ as follows,

\begin{equation}
S(P_{s,o})=\frac{1}{1+\sum_{i=2}^{n-1}\log d(o_i)}
\end{equation}
where $d(o_i)$ is the degree of entity $o_i$, i.e., the number of paths that entity $o$ participates. One approach is to optimize a path evaluation function: $\tau(c) = \max \mathcal{W}(P_{s,o})$, which maps the set of possible paths connecting $s$ and $o$ (i.e., $P_{s,o}$)to a truth value $\tau$. If $s$ is already present in the knowledge network, it can assign maximum truth value 1; otherwise the objective function will be optimized to find the shortest path between $s$ and $o$.

\subsubsection{Flow Optimization}
We can assume that each edge of the network is associated with two quantities: a \textit{capacity} to carry knowledge related to $(s, p, o)$ across its two endpoints, and a \textit{cost} of usage. The capacity can be computed using $S(P_{s,o})$, and the cost of an edge in knowledge is defined as $c_e = \log d(o_i)$. The goal is to identify the set of paths responsible for the maximum flow of knowledge between $s$ and $o$ at the minimum cost. The maximum knowledge a path $P_{s,o}$ can carry is the minimum knowledge of its edges, also called its bottleneck $B(P_{s,o})$. Thus, the objective can be defined as a minimum cost maximum flow problem,

\begin{equation}
\tau(e) = \sum_{P_{s,o}\in\mathcal{P}_{s,o}}B(P_{s,o})\cdot S(P_{s,o})
\end{equation}
where $B(P_{s,o})$ is denoted as a minimization form: $B(P_{s,o})=\min\{x_e|\in P_{s,o}\}$, with $x_e$ indicating the residual capacity of edge $x$ in a residual network~\cite{shiralkar2017finding}.

\textbf{Discussion} The knowledge network itself can be incomplete and noisy. For example, the entities in fake news claims may not correspond to any path exactly or may match multiple entities in the knowledge network. In this case, only performing path finding and flow optimization is not enough to obtain a good result to assess the truth value. Therefore, additional tasks (e.g.,  entity resolution, and link prediction) need to be considered in order to reconstruct the knowledge network and to facilitate its capability. Entity resolution is the process of finding related entities and creating links among them. Link prediction predicts the unseen links and relations among the entities.

\subsubsection{Stance Network Aggregation}
We can present the stance of users' posts either explicitly or implicitly. Explicit stances are direct expressions of emotion or opinion, such as Facebook's ``like'' actions. Implicit stances can be automatically extracted from social media posts.

% Explicit stance
Consider the scenario where the stances are \textit{explicitly} expressed in ``like'' actions on social media. Given the stance network $G_S=\{U\cup S\cup V, E_S\}$, the first step is to construct a bipartite graph $(U\cup V, L)$, where $L$ is the set of likes actions. The idea is that user express ``like'' actions due to both the user reputations and news qualities. The users and news items can be characterized by the Beta distributions $\text{Beta}(\alpha_i,\beta_i)$ and $\text{Beta}(\alpha_j,\beta_j)$, respectively. The distribution of a user $\text{Beta}(\alpha_i,\beta_i)$ represents the reputation or reliability of user $u_i$, and the distribution of a new piece $\text{Beta}(\alpha_j,\beta_j)$ represents the veracity of news $v_j$. The expectation values of the Beta distribution are used to estimate the degree of user reputation ($p_i=\frac{\alpha_i}{\alpha_i+\beta_i}$) or new veracity ($p_j=\frac{\alpha_j}{\alpha_j+\beta_j}$).  To predict whether a piece of news is fake or not, the linear transformation of $p_j$ is computed: $q_j=2p_j-1=\frac{\alpha_j-\beta_j}{\alpha_j+\beta_j}$, where a positive value indicates true news; otherwise it's fake news.

The model is trained in a semi-supervised manner. Let the training set consists of two subsets $V_F, V_T \subseteq V$ for labeled fake and true news, and $\Phi_i = \{u_i|(u_i,v_j)\in L\}$ and $\Phi_j = \{v_j|(u_i,v_j)\in L\}$. The labels are set as $q_j=-1$ for all $v_j\in I_F$, and $q_j=1$ for all $v_j\in I_T$, and q$q_j=0$ for unlabeled news pieces. The parameter optimization of user $u_i$ is performed iteratively by following updating functions:

\begin{equation}
\begin{aligned}
\alpha_i =& \Delta_{\alpha} + \sum_{q_i>0,i\in\Phi_i} q_i\\
\beta_i = &\Delta_{\beta} - \sum_{q_i<0,i\in\Phi_i} q_i\\
q_i = &(\alpha_i-\beta_i)/(\alpha_i+\beta_i)
\end{aligned}
\end{equation}
where $\Delta_{\alpha}$ and $\Delta_{\beta}$ are the prior base constants indicating the degree the user believing the fake or true news. Similarly, the parameter of news $v_j$ is updated as,

\begin{equation}
\begin{aligned}
\alpha_j =& \Delta_{\alpha}' + \sum_{q_j>0,j\in\Phi_j} q_j\\
\beta_j = &\Delta_{\beta}' - \sum_{q_j<0,j\in\Phi_j} q_j\\
q_j = &(\alpha_j-\beta_j)/(\alpha_j+\beta_j)
\end{aligned}
\end{equation}
where $\Delta_{\alpha}'$ and $\Delta_{\beta}'$ are the prior constants indicating the ratio of fake or true news. In this way, the stance (like) information are aggregated to optimize the parameters, which can be used to predict the news veracity using $q_j$~\cite{tacchini2017some}.

% Implicit stance
We can infer the \textit{implicit} stance values from social media posts, which usually requires a labeled stance dataset to train a supervised model. The inferred stance scores then serve as the input to perform fake news classification. Fake news pieces are more likely to receive controversial stances. Existing work mainly focuses on extracting hand-crafted linguistic features and using deep neural networks to obtain latent features, while few papers focus on network perspective. 

\section{Fake News Mitigation}~\label{sec:mitigation}

Fake news mitigation aims to reduce the negative effects brought by fake news. From a network analysis perspective, the goal is to minimize the scope of fake news spreading on social media. To achieve this, key spreaders of fake news need to be discovered such as provenances and persuaders . In addition, estimating the potential population affected by a fake news is useful for decision-makers to mitigate otherwise influential fake news. Moreover, choosing specific users to block the cascade of fake news, and even to start mitigation campaigns to immunize users are required to minimize the influence of fake news. 

\subsection{User Identification}
Identifying key users on social media is important to mitigate the effect of fake news. For example, the provenances of fake news indicates the sources or originators. Provenances can help answer questions such as whether the piece of news has been modified during its propagation, and how an ``owner'' of the piece of  information is connected to the transmission of the statement. In addition, it's necessary to identify influential persuaders to limit the spread scope of fake news by blocking the information flow from them to their followers on social media.

\subsubsection{Identifying Provenances}
% In social media, the provenance of fake news indicates the sources or originators. Provenances can help answer important questions such as whether the piece of news was modified during its propagation, and how an ``owner'' of the piece of  information is connected to the transmission of the statement. Searching provenance in social media is a new problem because social media lacks a centralized authority or mechanism that can store and certify provenance of a piece of social media data.  Hence, information provenance in social media is critical in understanding and identification of fake news. 

Given the diffusion network $G_D$, $P \subseteq U$, and a positive integer constant, $k \in Z^+$,  we identify the sources, $S \subseteq U$, such that $|S| \le k$, and $U(S, P)$ is maximized. Function $U(S, P)$ estimates the utility of information propagation starting from $S$ and ending at $P$. The Information Provenance (IP) problem estimates sources $\hat{S}$.
\begin{equation}
\hat{S} = \argmax_{S \in U, |S| \le k } U(S, P),
\label{eqn:one}
\end{equation}
$K\in Z^{+}$ is a positive integer because information in social media could originate from multiple sources. The solution to the above problem is based on the possible paths of information propagation using provenance seeking nodes, referred to as {\em provenance paths}. The provenance paths of information are usually unknown. In fact, the provenance paths for fake news dissemination in social media still remains an open problem. Here, we introduce how previous work define and approach this problem, which has potential to be applied and adapted to fake news research. The provenance paths objective can be formulated as,

\begin{equation}
\hat{G}^k = \argmax_{k \in Z^+} U(G^k),
\label{eqn:utility1}
\end{equation}
where $U(G^k)$ estimates the utility of the provenance paths $G^k$, for given $P$ nodes. The utility function depends on the underlying information propagation model. For example, as the independent cascade model, $U(G^k)$ is estimated as the product of all propagation probabilities, i.e., $U(G^k)= \prod_{(u \rightarrow v) \in G^k } p(u \rightarrow v)$. For a given graph $G$, there are exponentially many subgraphs having $k$-most roots and covering all the nodes in $P$. The problem aims to extract a subgraph ($\hat{G}^k$) with the maximum utility. This problem is NP-complete, and challenges arise because: 1) only a few nodes ($P$) are given, and 2) $G_D$ is usually in large-scale.

To solve this problem, existing work considers the node properties, such as node centralities, to capture the potential provenance nodes. Some observations are commonly revealed in information diffusion process: (1)  {\em Degree propensity} suggests that the higher degree centrality nodes in a network are more likely to be transmitters than the randomly selected nodes, and (2) {\em Closeness propensity} reveals that the higher degree nodes closer to the nodes in $P$ are more likely to be transmitters than the randomly selected higher degree nodes. Based on the Degree Propensity and Closeness Propensity hypotheses, the nodes with higher degree centralities that are closer to the P-nodes are more likely to be transmitters. Hence, as shown in~\cite{barbier2013provenance}, the method can estimate top $m$ transmitters, from the given set of $P$ nodes. Let M be a set of these top $m$ transmitters, then the provenance paths can be recovered using an approximate algorithm as in~\cite{barbier2013provenance}. The basic idea and procedure is as follows in Algorithm~\ref{alg:find_path}.

\begin{algorithm}
\caption{Finding Provenance Paths}\label{alg:find_path}
\begin{algorithmic}[1]
\Require{$G_D$, $P$, Transmitters $M$, $k$}
\Ensure{$G^k$, S}
\State $G^k \leftarrow \cup_{c\in M} \text{dst} (G_D,c,P)$ \Comment{Combining all the shortest paths from a node $c$ to each of the $P$ nodes}
\State $S\leftarrow \text{find\_sources} (G^k)$ \Comment{Initializing sources from the roots of $G^k$}
\While {$|S|\geq k$}
\State $c\leftarrow \text{MinComAnc} (G_D,S)$ \Comment{Finding a common node  with average distance from node c to any pair of nodes $(u, v) \in S$ is the minimum}
\State $G^k\leftarrow G^k\cup \text{dst}(G_D,c,S)$
\State $S\leftarrow \text{find\_sources} (G^k)$
\If{$|S|$ \text{not decreasing} }
\State$S\leftarrow \text{GreedSel}(G_D,S,P,k)$ \Comment{Greedily selecting k nodes in $S$ such that each $s \in S$ can reach the maximum number of $P$ nods}
\EndIf
\EndWhile
\end{algorithmic}
\end{algorithm}

\subsubsection{Identifying Persuaders}
% To minimize the impact of fake news, it's necessary to identify these influential persuaders. Agarwal \textit{et al.}~\cite{agarwal2008identifying} study a related problem of identifying influential bloggers on a blog site with abundant content and statistic features. 
Leadership-theory suggests that opinion leaders likely to influence their followers' actions and beliefs~\cite{kirkpatick1991leadership}. Given this, a leadership-based approach should be able to identify $K$-leaders such that they perform actions on the maximum number of posts. Then, given the diffusion network $G_D$, the first step is to construct a bipartite network $G_D'=(U\cup S, E)$, with $U$ being the set of users and $S$ being the posts they write. The problem of identifying K leaders is similar to the minimum dominating set problem, which is NP-complete even for bipartite graphs. Therefore, the cost of selecting user $u_i$ as a leader is equivalent to the total number of posts she acted on, i.e, $c(\{u_i\}) = \mathbf{1}_{e(u_i, s_j) \in E}$ where $\mathbf{1}$ is the indicator function. Similarly, the cost of selecting set $U'$ as leaders can be computed as $c({U'}) = \sum_{u_i\in U'}\mathbf{1}_{e(u_i, s_j) \in E}$. The cost function $c:\mathbb{P}(U') \rightarrow \mathbb{R}$, where $\mathbb{P}(U')$ denotes the power set of users ${U'}$, which satisfies the following properties: (1) {\em non-negative:} for every ${U'} \subseteq {U}$, $c({ U'}) \ge 0$; (2) {\em non-decreasing:} for every ${U'_1}, {U'_2} \subseteq {U}$ with ${U'_1} \subseteq {U'_2}$, $c({U'_1}) \le c({U'_2})$; and (3) {\em sub-modular:} for every ${U'_1}, {U'_2} \subseteq {U}$ with ${U'_1} \subseteq {U'_2}$ and every $u \in {U} \setminus {{U'_1}}$, $c({{U'_1}} \cup \{u\}) - c(\{u\}) \ge c({{U'_2}} \cup \{u\}) - c(\{U'_2\})$. Based on the properties, the problem of finding $K$-leaders is tantamount to the maximization of a non-negative, non-decreasing, sub-modular function with cardinality constraint. A hill-climbing algorithm can solve this problem with a provable constant approximation.

\subsection{Network Size Estimation}

The fake news diffusion process has different stages in terms of people's attention and reactions over time, resulting in a unique life cycle different from that of in-depth news~\cite{castillo2014characterizing}.  The impact of fake news on social media can be estimated as the number of users that are potentially affected by the news piece, an amount we want to assess and then minimize. We can adapt the network scale-up based method~\cite{bernard1990comparing} for estimating the size of uncountable populations to estimate the size of the population affected by fake news on the provenance paths discussed previously.
%We can estimate personal network sizes using the network scale-up methods~\cite{bernard1990comparing}. Based on findings from the applications of this method, the  population affected by fake news can be further approximated based on the provenance paths discussed previously. 
Specifically, we can assume that two different approaches $\Omega_A$ and $\Omega_B$ are utilized to find provenance paths, and the resultant nodes that in the paths are the receivers of the fake news, i.e., $R_A$ and $R_B$. If we assume that the two methods produce independent results, then the quantity $|R_A \cap R_A|/ |R_A|$ is an estimate of the fraction of the recipients covered by method $A$. Then the scale-up based method suggests that the total number of recipient $N$ can be estimated as,
\begin{align}
N = R_A * \frac{|R_B|}{|R_A \cap R_B|}
\end{align} 
However, the two methods $\Omega_A$ and $\Omega_B$ may not produce independent results as social media data is intrinsically linked. For example, $\Omega_A$ and $\Omega_B$ can be the results from different platforms (e.g., Facebook and Twitter), thus $|R_A \cap R_B|$ can be better estimated as $R_a$ and $R_b$ on two platforms that are likely more independent than $R_A$ and $R_B$ on the same platform. Various methods of linking user identities~\cite{shu2017user} can be applied to identify the overlaps of $R_A$ and $R_B$.

\subsection{Network Intervention}

% \textbf{\color{red}Careful verification is needed***}

The goal of network intervention is to develop strategies to control the widespread dissemination of fake news before it goes viral. Network intervention mainly consists of two perspectives as follows: 

\subsubsection{Influence Minimization}

Limiting the spread of fake news can be seen as analogous to inoculation in the face of an epidemic. Models of epidemics generally assume that a global parameter describes the probability that a user is infected by a neighbor. This assumption is violated in real-world situations of information exchange where users have varying degrees of willingness to accept information from their neighbors. Thus, the Independent Cascade Model (ICM)~\cite{saito2008prediction} is proposed to alleviate this problem by assuming each edge has its specific activation probability. ICM is denoted as a sender-centric model. Specifically, the node that becomes active at time $i$ has, in the next time step $i+1$, one chance of activating each of its neighbors. Let $u$ be an active node at time $t$. Then, for any neighbor $u'$, there is a probability $p_{u,u'}$ that node $u'$ gets activated at $i+1$. Our goal is to stop the information cascade and thus to limit the influence of fake news. This can be treated as an influence minimization problem: Given the diffusion network $G_D$ with the initial infected user set $U'\subseteq U$, the goal is to minimize the number of final infected users by blocking $k$ users of the set $D\subseteq U$ such that the following function is optimized,

\begin{equation}
D^* = \text{argmin}~\sigma(S| {U}\setminus{U'})
\end{equation}
where $\sigma(S| {U}\setminus{U'})$ denotes the influence of set $U'$ when nodes in the set $D$ are blocked. The intuitive criteria to obtain the ``blocking user'' set $D$ include: i) limiting the number of out-links of the sender node and potentially reducing the chance of activating others; ii) limiting the number of in-links of receiver nodes and therefore reducing their chance of being activated by others; and iii) decreasing the activation probability of a node ($p_{uu'}$) and therefore reducing the chance of activating others.

\subsubsection{Mitigation Campaign}
Limiting the impact of fake news is not only to minimize the spread of fake news but also maximize the spread of true news.  The campaign to mitigate fake news and to maximize true news forms during the information diffusion process. The network activities of fake news and real news can be represented as Multivariate Hawks Processes (MHP) with self and mutual excitations, where the control incentivizes more spontaneous mitigation events~\cite{farajtabar2017fake}. The influence of fake and real news is quantified using event exposure counts, represented by the number of times users are exposed to the news. The goal is to optimize the activity policy of a set of campaigner users to mitigate a fake news process stemming from another set of users. The whole idea is to optimize the performance of real news propagation (through the campaigner users) in diffusion network, ensuring that people who are exposed to fake news are also exposed to real news, so that they are less likely to be convinced by fake news. Specifically, we define fake news MDP as $F(t)=(F_1(t),F_2(t),...,F_m(t))$, where $F_i(t)$ is the number of times that user $u_i$ shares a pieces of news from fake news campaign up to time $t$; similarly $M(t)$ for mitigation campaign. The objective is to maximize the common exposure of fake and real news, with a reward function $R$ in the current stage $t$,

\begin{equation}
R(x^k,u^k) = \frac{1}{m}\mathcal{F}^t(\tau_{t+1};x^t,u^t)\mathcal{M}^t(\tau_{t+1};x^t,u^t)
\end{equation}
where $\mathcal{M}(\cdot)$ ($\mathcal{F}(\cdot)$) is the exposure process for news in mitigation (fake) campaign. The friendship network $G_F$ can be represented by a user-user adjacency matrix $A\in\{0,1\}^{m\times m}$,  we have $\mathcal{F}^t=AF(t)$ and $\mathcal{M}^t=AM(t)$. $x^t$ is the state representation, and $u^t$ is the intensity value, and $\tau_{t+1}$ is the time of next stage.
\begin{figure}[htbp!]
   \centering
   \includegraphics[scale=0.47]{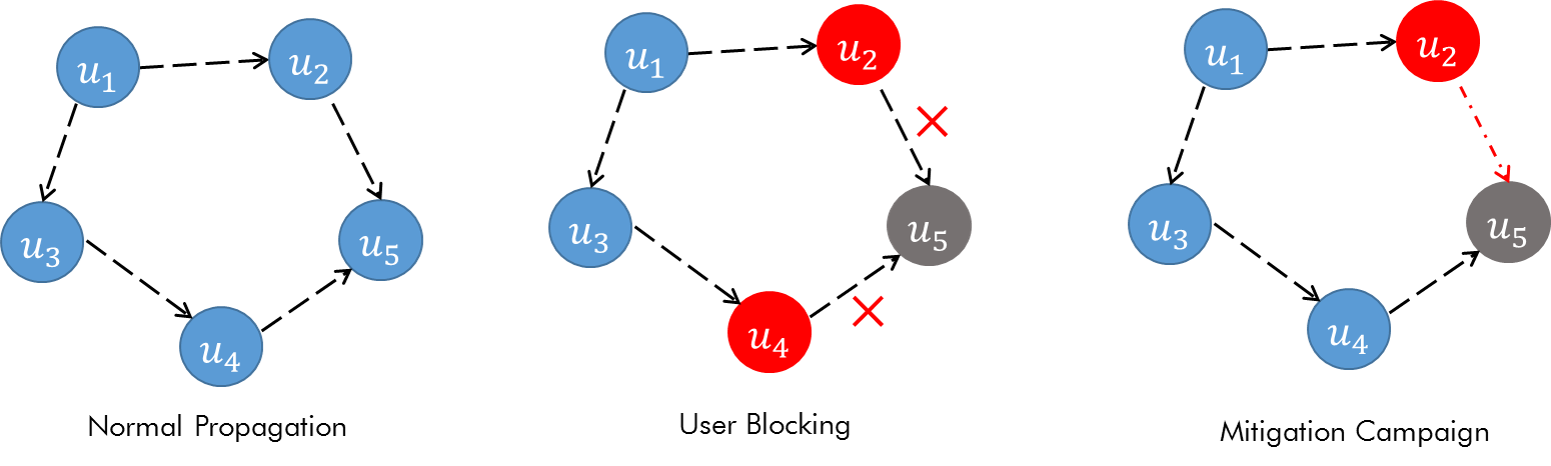}
   \caption{The Influence limitation strategies: user blocking and mitigation campaign.}
   \label{fig:influence_mini}
\end{figure}

As shown in Figure~\ref{fig:influence_mini}, the ``user blocking'' strategy assumes that each user can only be in one state (i.e., normal, infected, and blocked), and the state will not change. For example, since user $u_2$ and $u_4$ are blocked, $u_5$ will not be infected. As tested in~\cite{farajtabar2017fake}, the ``mitigation campaign'' strategy assumes people exposed more to fake news should also be exposed more to true news, so they are less likely to believe completely in fake news. For example, if user $u_5$ receives both information from the fake campaign through user $u_4$ and from the mitigation campaign through $u_2$, he/she is less likely to believe the fake campaign in the case of a user who does not receive the mitigation campaign.
%% Need mote explanations and consistency

\section{Summary} \label{sec:discuss}
% Try to put all previous suff into a big picture 
This chapter presents some recent trends in studying fake news on social media via network analysis. During fake news dissemination, different entities are involved that can be categorized into content, social and temporal dimensions. In addition, the inherent network properties motivate and strengthen the needs to perform network analysis to study fake news. The dimensions of fake news dissemination reveal mutual relations and dependencies that can form different types of networks.  Based on these networks, we introduce some representative methods to demonstrate how to perform fake news detection and mitigation.

\section*{Acknowledgements}

This material is based upon work supported by, or in part by, the ONR grant N00014-16-1-2257, N000141310835, and N00014-17-1-2605.

\bibliographystyle{splncs03}
{\bibliography{paper}}

\end{document}